\documentclass[12pt]{iopart}

\usepackage{iopams}
\usepackage[english]{babel}
\usepackage{graphicx}
\usepackage{cite}
\usepackage{bbm}
\usepackage{mathenv}
\usepackage{color}
\usepackage{float}
\usepackage{braket}
\begin{document}

\title{Entropy fluctuations as a mixedness quantifier}

\author{Jorge A. Anaya-Contreras,$^1$ Arturo Z\'u\~niga-Segundo$^1$ and H\'ector M. Moya-Cessa$^2$}

\address{$^1$Instituto Polit\'ecnico Nacional, ESFM, Departamento de F\'isica. Edificio 9, Unidad Profesional ``Adolfo L\'opez Mateos,'' CP 07738 CDMX, Mexico\\ $^2$Instituto Nacional de Astrof\'{\i}sica, \'Optica y Electr\'onica INAOE\\ Calle Luis Enrique Erro 1, Santa Mar\'{\i}a Tonantzintla, Puebla, 72840 Mexico}

\begin{abstract}
We propose a mixedness quantifier based on entropy fluctuations.  It provides information about the degree of mixedness either for finite dimensional and infinite dimensional Hilbert spaces. It may be used to determine the reduction of the Hilbert space as it becomes maximum when, either the state is maximally mixed, or, when the Hilbert space  effectively reduces its dimensions, such as in the atom field interaction where the two-level atom dictates the final dimension of the field.
\end{abstract}



\maketitle
\section{Introduction}

 {It is well known that the resources provided by quantum coherence are severely affected by an environment. A system-environment interaction produces a loss of information in the quantum system, measured in terms of the loss of its purity. It has been shown that there are limits imposed  by the degree of the mixedness of a quantum system on the amount of quantum coherences that it may have, in fact, it has been demonstrated that mixedness and quantum coherences satisfy a complementarity relation that is crucial in the understanding the interplay between quantum resources and the effect of environments on them.  \cite{Singh}}.  

 Phenomena such as decoherence or entanglement  \cite{Barnett} can easily destroy the purity of a system initially described by a wavefunction and take it  to an statistical mixture of states. Decoherence, that may be produced by the interaction of a given system with its environment (a big enemy of non-classical states) affects pure states \cite{Expla} in such a way that produces such statistical mixtures. On the other hand, entanglement between two systems, when they are separated and no conditional measurements are realized on them -for instance an atom going through a cavity, gets entangled with the field inside, and if it exits the cavity without being measured- produces loss of information reflected in the fact that the field density matrix becomes, in general, an statistical mixture of field states.
  {Studies of mixedness and coherence have been proposed in several systems such as photon polarization and meson and neutrinos \cite{Dixit,Kwait2005,Peters2004,Peters2004b} and measurements of quantum correlations in mixed state metrology have been put forward \cite{Kahlid}.}
 
 In order to know how pure is a given density matrix, there are  good measures of purity, namely the linear entropy or the von Neumann entropy \cite{Neumann}. However, the degree of mixedness, that would reflect, among other things, how strong was the interaction with the environment or how entangled were the atom and the field, is a parameter not well determined yet.

Probably the most common tools to assess the degree of mixedness of a given quantum  state are entropy \cite{Barnett,Neumann,Pathak} and the so-called linear entropy \cite{Phoenix2,Phoenix3}. They define properly the purity of a given state and, because they depend on the dimensionality of the Hilbert space, the degree of mixedness has to be defined for each problem. This is due to the fact that for some specific density matrices, it is needed to have {\it a priori} information about which states conform it, for instance if it is a mixture of several coherent states, we need to know the amount of such states that give rise to such the density matrix.  {This just means that for infinite Hilbert spaces such as the ones required for quantized fields, a normalization is not possible, and, only when the Hilbert space is effectively reduced, as in the case when it interacts with an spin system, the final dimensionality of the system has to be known in order to properly normalize the entropy.  }

It is therefore clear that the von Neumann entropy is a good measure of purity, but, there is still no good measure of the degree of mixedness  {as it depends on the dimension of the Hilbert space (and for infinite Hilbert spaces there is not a correct way of normalization). The main interest in the present contribution is to study the case of infinite Hilbert spaces such as the ones that describe quantized fields, as for this specific states a measurement of mixedness is not yet well understood. For instance a measurement of entropy of, say, $\ln 2$ has different meaning for a two level system such as an statistical mixture of two large coherent states, or for an statistical mixture of $N\ge 3$ coherent states, or a more extreme case such as a thermal state. In next section we introduce a parameter that tell us either the degree of mixedness in an infinite system or if it has effectively evolved into a finite dimensional state, namely by its interaction of it with a second system (we study in particular the atom-field interaction) or because of its interaction with an environment. In order to have an initial test for the parameter, we apply it to a two-level spin system and then to a quantized field described by several states, namely, statistical mixtures of coherent states and a thermal distribution of states. In order to present the mixedness parameter (MP), we introduce, still in this Section, the tools commonly used to work with degrees of mixedness or purity such as entropy and linear entropy.} 
\subsection{Entropy} 
The quantum mechanical entropy is defined as \cite{Neumann}
\begin{equation}
S=\langle \hat{S} \rangle = \langle - \ln\hat{\rho} \rangle
=-Tr\{\hat{\rho}\ln\hat{\rho}\},
\end{equation} 
and is also known as the von Neumann entropy. It delivers information about the purity of a given state $\rho$. It may be seen as the expectation value of the entropy operator \cite{Moya} $\hat{S}=- \ln\hat{\rho}$.

Depending on the density matrix state, we have that for a pure state,  $S=0$,
while if it is in a mixed state, $S>0$. This makes $S$ a good measure of 
the deviation from  pure states.  Because the density matrix of the system,
$\rho(t)$, is governed by a unitary time evolution operator, the entropy of a {\it closed} system is time independent.

But we usually do not have closed systems, as systems may
interact with other systems  and/or with an environment, making the entropy to evolve during those kind of interactions. If we consider a system
composed by two sub-systems, although the entropy of the whole system does not change in time, we can ask ourselves about the entropy of each 
subsystem. If we call one sub-system $A$ and the other $B$, then
the trace of the total density matrix on the $A$ subsystem basis
gives us the density matrix for the $B$ subsystem
\begin{equation}
\hat{\rho}_B=Tr_A\{\hat{\rho}\},
\end{equation}
and viceversa
\begin{equation}
\hat{\rho}_A=Tr_B\{\hat{\rho}\}.
\end{equation}
The entropies for $A$ and $B$ may be defined as
\begin{equation}
S(\hat{\rho}_{A,B})=-Tr_{A,B}\{\hat{\rho}_{A,B}\ln\hat{\rho}_{A,B}\}.
\end{equation}
The effect of tracing over one of the subsystems variables means
that each subsystem is no longer governed by a unitary time
evolution, which produces that the entropy of  each subsystem
becomes time dependent and it may evolve now from  pure states to
 mixed states (or viceversa).

Araki and Lieb \cite{Araki} stated  the following
inequalities for two interacting subsystems
\begin{equation}
|S(\hat{\rho}_{A})-S(\hat{\rho}_{B})|\le S \le S(\hat{\rho}_{A})+S(\hat{\rho}_{B}). \label{araki}
\end{equation}
Therefore, if the two subsystems are initially in a pure state,
the whole entropy is zero ($S=0$), such that  both subsystems will have the same entropy,
$S(\hat{\rho}_{A})=S(\hat{\rho}_{B})$, in such a way that if the Hilbert space of one subsystems is smaller than the other, it will dictate the maximum entropy of the large one.
\subsection{Linear entropy}
Another common tool to study the {\it purity} of  a state is by
means of the so-called linear entropy, $\xi= \langle \hat{\xi} \rangle
$,
\begin{equation}
\xi= \langle (1-\hat{\rho}) \rangle =1-Tr\{\hat{\rho}^2\}.
\end{equation}
By using the eigenbasis of the density matrix it can be shown that
\begin{equation}
Tr\{\hat{\rho}^2\}=\sum_{n}\rho_n^2\le \sum_{n}\rho_n=1.
\end{equation}
Because the equality holds only for pure states, the purity parameter, $\xi$
discriminates uniquely between mixed and pure states. By using the
fact that $1-\rho_n\le -\ln\rho_n$ for $0<\rho_n\le 1$ a lower  bound for the entropy is found
\begin{equation}
\xi\le S.
\end{equation}

\section{Mixedness parameter}
We now introduce a MP based on entropy fluctuations
\begin{equation}
(\Delta S)^2=\langle \hat{S}^2 \rangle - \langle \hat{S} \rangle^2,
\end{equation} 
as follows
\begin{equation}
Q_S=\exp\left[-\frac{(\Delta S)^2}{S}\right].
\end{equation} 
 {From an statistical point of view, it is important to study dispersion measurements of different observables. For instance, in the case of the average number of photons, we may use the Mandel-$Q$ parameter \cite{Mandel},
\begin{equation}
Q_M=1-\frac{(\Delta \hat{n})^2}{\langle \hat{n}\rangle},
\end{equation} 
as a measurement of the degree of subpoissonicity of a given state (where $\hat{n}$ is the so-called number operator). This motivates us to use the entropy operator, $\hat{S}=- \ln\hat{\rho}$, as an adequate measurement of mixedness of a state, as it is not only bounded, but also does not require {\it a priori} information of the density matrix. }

 {The quantity $(\Delta S)^2=\langle \hat{S}^2 \rangle - \langle \hat{S} \rangle^2,$, from an statistical point of view would represent how disperse is the statistical mixture in reference to the entropy value, therefore, we may say that $Q_S$ contains the same amount of information as von Neumann and linear entropies do, but with the profit that there is no need of normalization such that we may apply it to quantized fields that live in infinite Hilbert spaces.}

The parameter $Q_S$ is bounded from zero for pure states to one for either completely mixed states or a reduction of the Hilbert space to an effective lower dimension one.  {The MP is independent of the dimension of the Hilbert space, although it may reflect its effective reduction to a lower dimensional one, even from infinite dimensional to, effectively, finite).}

 {We show that it may be applied to time dependent density operators such as the Jaynes-Cummings model  \cite{JCM}.}
\section{Degree of mixedness for several states}

\subsection{Two-level system} 
Consider the state of a two-level system given by the density matrix
\begin{equation}
\label{T33}
\hat{\rho}=\cos^{2}\phi\ket{e}\bra{e}+\sin^{2}\phi\ket{g}\bra{g}\,,
\end{equation}
it is direct to show that the entropy is
\begin{equation}
\label{T34}
S=-\cos^{2}\phi\ln\left(\cos^{2}\phi\right)-\sin^{2}\phi\ln\left(\sin^{2}\phi\right)\,,
\end{equation}
while entropy fluctuations may be easily calculated as
\begin{equation}
\label{T35}
\Delta S =\frac{1}{2}\left|\sin2\phi \ln(\cot^{2}\phi)\right|\,.
\end{equation}
In Figure 1 we plot the normalized entropy, $S/\ln 2$, and the quantifier we are introducing to measure mixedness, the MP. It may be observed a similar behaviour.
\begin{figure}[h!]
\centering
\includegraphics[width=12cm]{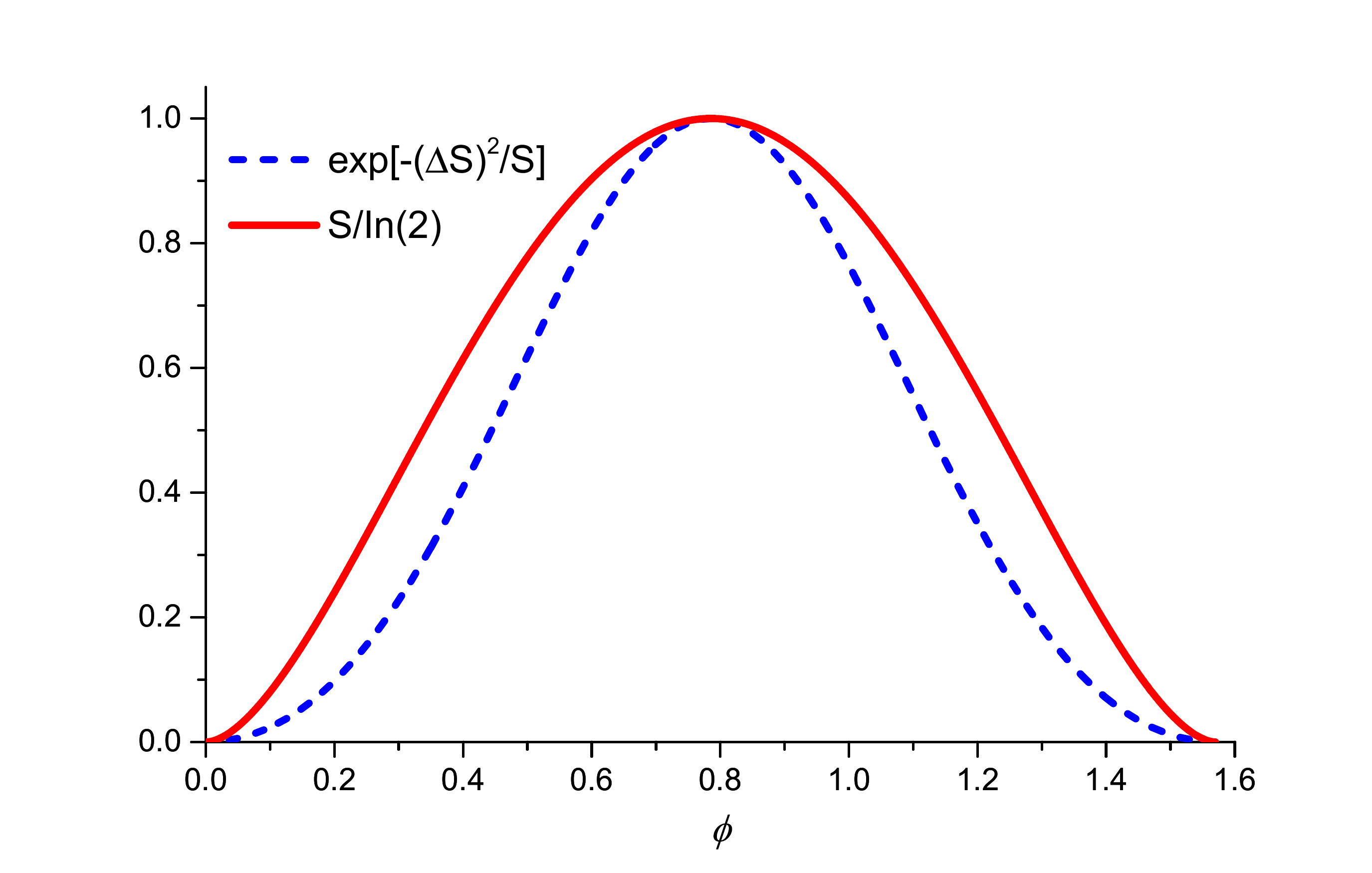} 
\caption{We plot the normalized entropy, $S/\ln 2$, and   the MP $e^{-(\Delta S)^{2}/S}$  for the two-states density matrix $\hat{\rho}=\cos^{2}\phi\ket{e}\bra{e}+\sin^{2}\phi\ket{g}\bra{g}$ as a function of $\phi$.}
\label{SDSatom}
\end{figure}

 {\subsection{Systems living in an infinite Hilbert space}}
In order to show that this parameter may be applied to  {systems living in  infinite Hilbert spaces, and as examples}, we use it here for three different cases of  quantized fields:  statistical mixtures of two and three coherent states and a thermal field.
 
$\textbf{I)}$ An statistical mixture of two coherent states \cite{Glauber} is written as 
\begin{equation}
\label{T26}
\hat{\rho}=\frac{1}{2}\ket{\alpha}\bra{\alpha}+\frac{1}{2}\ket{-\alpha}\bra{-\alpha}\,,
\end{equation}
where $|\alpha\rangle$ is a coherent state, may be purified \cite{Araki,Anaya} to the wave function
\begin{equation}
\label{T26.5}
|{\Psi}\rangle=|1\rangle\ket{\psi_{1}}+\ket{2}\ket{\psi_{2}}, 
\end{equation}
with  the unnormalized wavefunctions  $|\psi_1\rangle=\frac{1}{\sqrt{2}}\ket{\alpha}$ and $|\psi_2\rangle=\frac{1}{\sqrt{2}}\ket{-\alpha}$. The above expression allows to find the entropy of the state (\ref{T26}) in a simply way
\begin{equation}
\label{T27}
S=-\lambda_{1}\ln\lambda_{1}-\lambda_{2}\ln\lambda_{2}\,,
\end{equation}
as well as its entropy fluctuations
\begin{equation}
\label{T28}
\Delta S=\sqrt{\lambda_{1}\lambda_{2}}\left|\ln\left(\frac{\lambda_{1}}{\lambda_{2}}\right)\right|\,,
\end{equation}
where $\lambda_{1}$ and $\lambda_{2}$ are the eigenvalues of the matrix
\begin{equation}
\label{T29}
\left(\begin{array}{ll}
\braket{\psi_{1}|\psi_{1}}& \braket{\psi_{1}|\psi_{2}}^{*}\vspace{0.4cm}\\
\braket{\psi_{1}|\psi_{2}}&\braket{\psi_{2}|\psi_{2}}
\end{array}\right)\,.
\end{equation}
Equations (16) and (17) may be written explicitly as
\begin{equation}
\label{T31}
S=-\left(\frac{1+e^{-2|\alpha|^{2}}}{2}\right)\ln\left(\frac{1+e^{-2|\alpha|^{2}}}{2}\right)-\left(\frac{1-e^{-2|\alpha|^{2}}}{2}\right)\ln\left(\frac{1-e^{-2|\alpha|^{2}}}{2}\right)\,,
\end{equation}
and
\begin{equation}
\label{T32}
\Delta S = \frac{\sqrt{1-e^{-4|\alpha|^{2}}}}{2}\ln\left(\frac{1+e^{-2|\alpha|^{2}}}{1-e^{-2|\alpha|^{2}}}\right)\,.
\end{equation} 
In Figure 2 we plot the normalized entropy and the mixedness parameter based on entropy fluctuations. They show a similar behaviour. However, note that in order to normalize the entropy in this (infinite) case, we had to have knowledge that the field was an statistical mixture of {\it two} coherent states.  {In the case of the mixedness parameter, it show the effective reduction of the Hilbert space from an infinite  to a finite two-dimensional Hilbert space.}

 {It is well-known  that coherent states become orthogonal as they increase their amplitude ($\langle |\alpha|-\alpha\rangle| \approx 0 $ for $|\alpha|\gg 0$ ). This implies that the infinite Hilbert space where the density matrix, given in Equation (15), lives, effectively reduces its dimension to two possible states, generating a completely mixed state as $\alpha$ increases. The mixedness parameter provides such information without the need of knowing {\it a priori} the number of coherent states conforming the density matrix.}

$\textbf{II)}$ If the statistical mixture (\ref{T26}) is substituted by a mixture of three coherent states

\begin{figure}[h!]
\centering
\includegraphics[width=12cm]{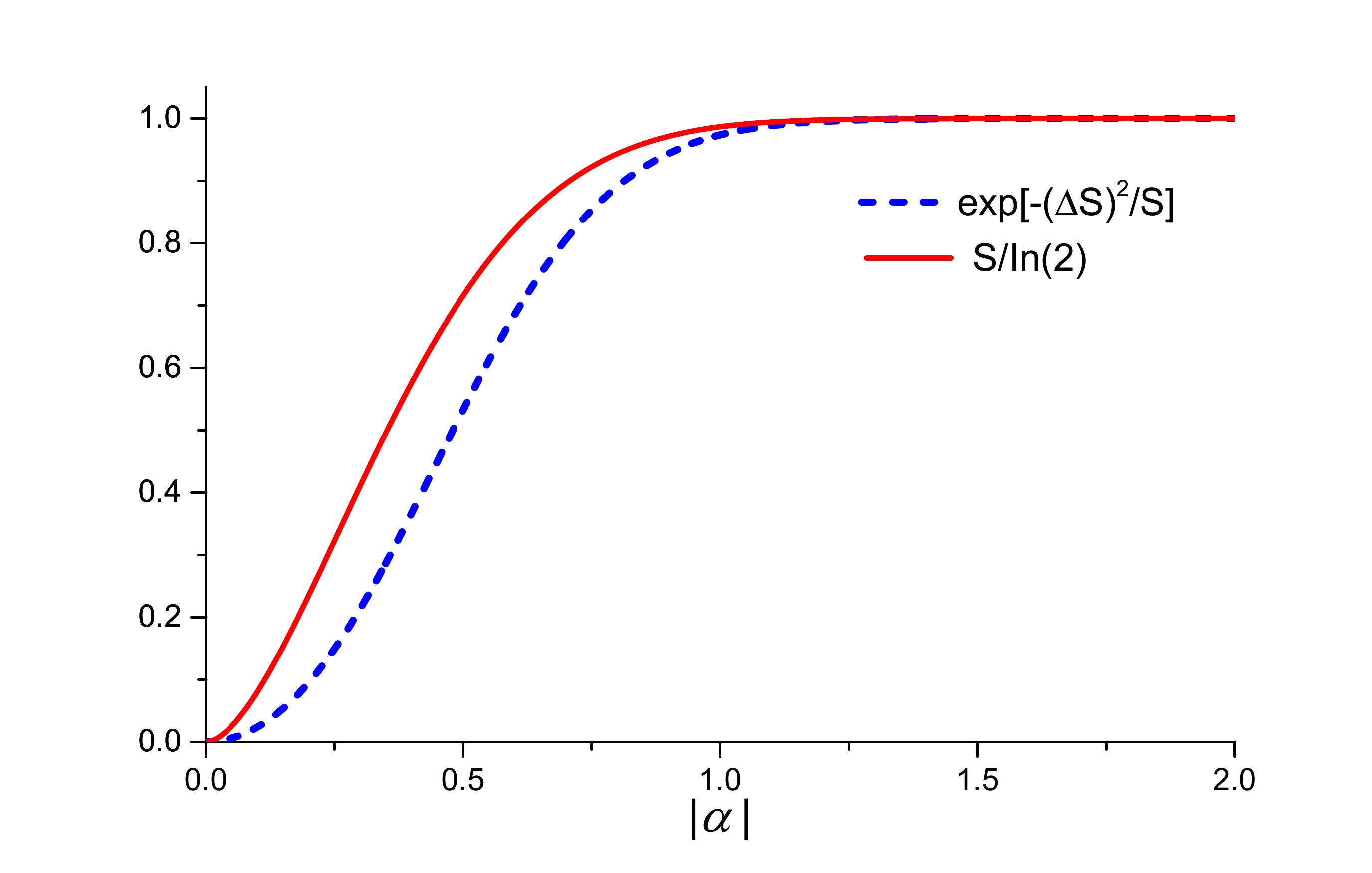} 
\caption{We plot the normalized entropy, $S/\ln2$, and  $e^{-(\Delta S)^{2}/S}$ for the field density operator $\hat{\rho}=(\ket{\alpha}\bra{\alpha}+\ket{-\alpha}\bra{-\alpha})/2$ as a function of $|\alpha|$.}
\label{SDScat}
\end{figure}

\begin{equation}
\label{T36}
\hat{\rho}=\frac{1}{3}\ket{\alpha}\bra{\alpha}+\frac{1}{3}\ket{-\alpha}\bra{-\alpha}+\frac{1}{3}\ket{2\alpha}\bra{2\alpha}\,,
\end{equation}
the entropy and the MP may be written, respectively as\begin{equation}
\label{T37}
S=-\lambda_{1}\ln\lambda_{1}-\lambda_{2}\ln\lambda_{2}-\lambda_{3}\ln\lambda_{3}\,,\qquad  \Delta S = \sqrt{\lambda_{1}(\ln\lambda_{1})^{2}+\lambda_{2}(\ln\lambda_{2})^{2}+\lambda_{3}(\ln\lambda_{3})^{2}-S^{2}}\,,
\end{equation}
where $\lambda_{1}$, $\lambda_{2}$ and $\lambda_{3}$ are the eigenvalues of the matrix
\begin{equation}
\label{T39}
\left(\begin{array}{lll}
\frac{1}{3}&\frac{1}{3}e^{-2|\alpha|^{2}}&\frac{1}{3}e^{-\frac{|\alpha|^{2}}{2}}\vspace{0.4cm}\\
\frac{1}{3}e^{-2|\alpha|^{2}}&\frac{1}{3}&\frac{1}{3}e^{-3|\alpha|^{2}}\vspace{0.4cm}\\
\frac{1}{3}e^{-\frac{|\alpha|^{2}}{2}}&\frac{1}{3}e^{-3|\alpha|^{2}}&\frac{1}{3}
\end{array}\right)\,.
\end{equation}
\begin{figure}[h!]
\centering
\includegraphics[width=12cm]{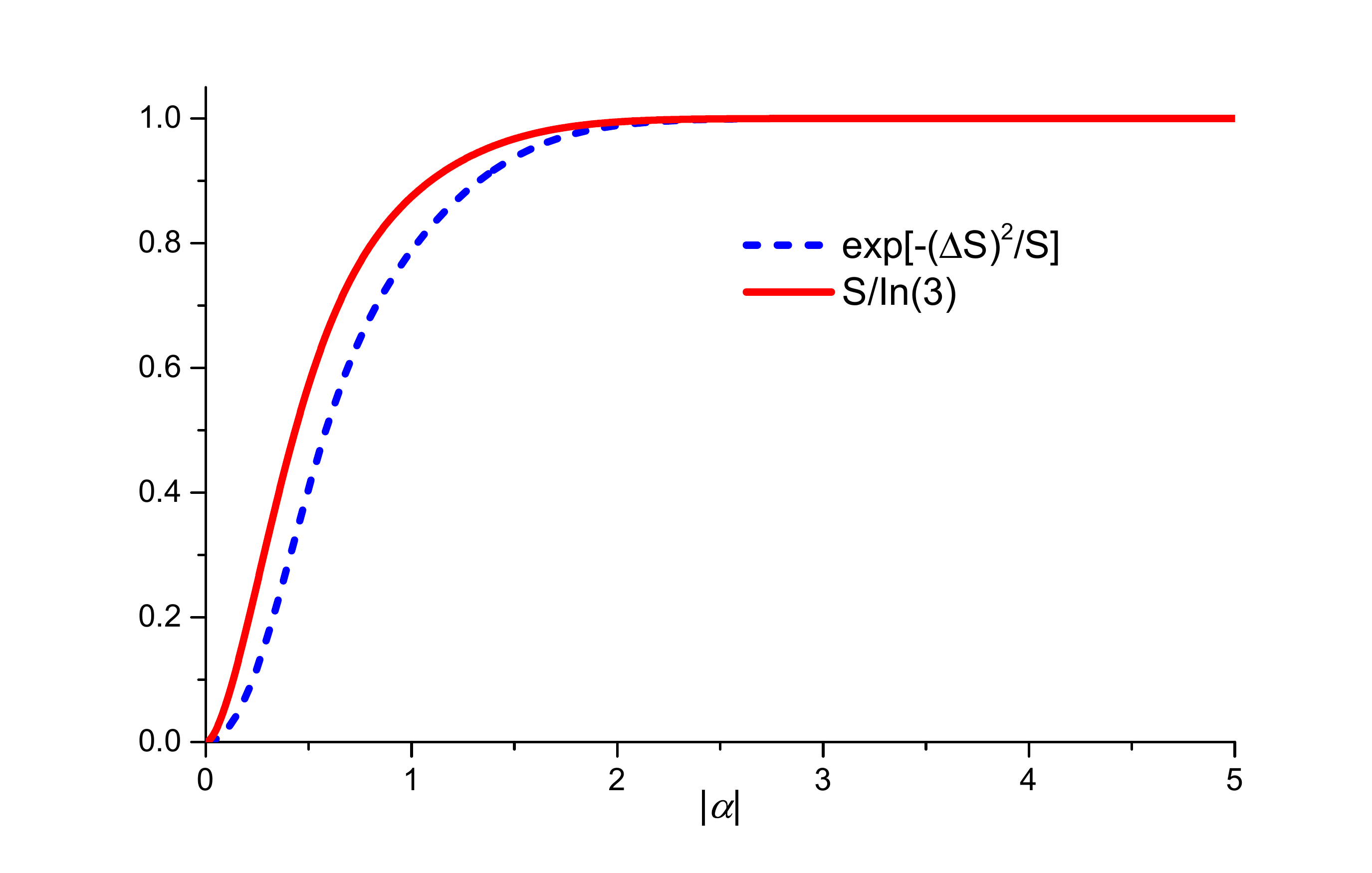} 
\caption{We plot the normalized entropy, $S/\ln3$ and the MP, $e^{-(\Delta S)^{2}/S}$ for the statistical mixture $\hat{\rho}=(\ket{\alpha}\bra{\alpha}+\ket{-\alpha}\bra{-\alpha}+\ket{2\alpha}\bra{2\alpha})/3$ as a function of $|\alpha|$.}
\label{SDSquad}
\end{figure}
In Figure 3 we plot the normalized entropy and the mixedness parameter based on entropy fluctuations. In this case, again we had to have {\it a priori} knowledge of the number of states producing the field density matrix in order to properly normalize the entropy. We note, again, similar behaviour for both parameters. 

 {This case shows us that the infinite Hilbert space where the density matrix given in Equation (22) lives, effectively reduces its dimension to three possible states, generating a {\it lower dimensional} mixed state as $\alpha$ increases. }

$\textbf{III)}$ The entropy and MP for a thermal distribution, 
\begin{equation}
\label{Th1}
\hat{\rho}=\sum_{n=0}^{\infty}P_{n}\ket{n}\bra{n}\,,
\end{equation}
where $\left\lbrace\ket{n}\right\rbrace_{n=0,1,2,...}$ are number states and  $P_{n}$ is defined in terms of the average number of thermal photons,  $\bar{n}$, as $\displaystyle{P_{n}=\frac{\bar{n}^{n}}{(1+\bar{n})^{n+1}}}$, may be easily found (because the density matrix is diagonal in the Fock basis). They are
\begin{equation}
\label{Th2}
S=\sum_{n=0}^{\infty}P_{n}\ln P_{n}\,, \qquad \Delta S =\sqrt{\sum_{n=0}^{\infty}P_{n}\ln^{2}P_{n}-S^{2}}\,.
\end{equation}
In this case it is not clear how the entropy may be normalized. In Figure 4 we plot both quantities as a function of the average number of photons. It may be clearly seen how the MP is bounded while the entropy is not.

 { The increase in the mixedness parameter is because it passes form a pure state ($\bar{n}=0$)  to a mixed state ($\bar{n}\ne 0$). Although a thermal state is an state of maximum entropy, to be completely mixed would require all the coefficients in Equation (25) to be equal, which is unphysical as would require infinite energy.}

\begin{figure}[h!]
\centering
\includegraphics[width=12cm]{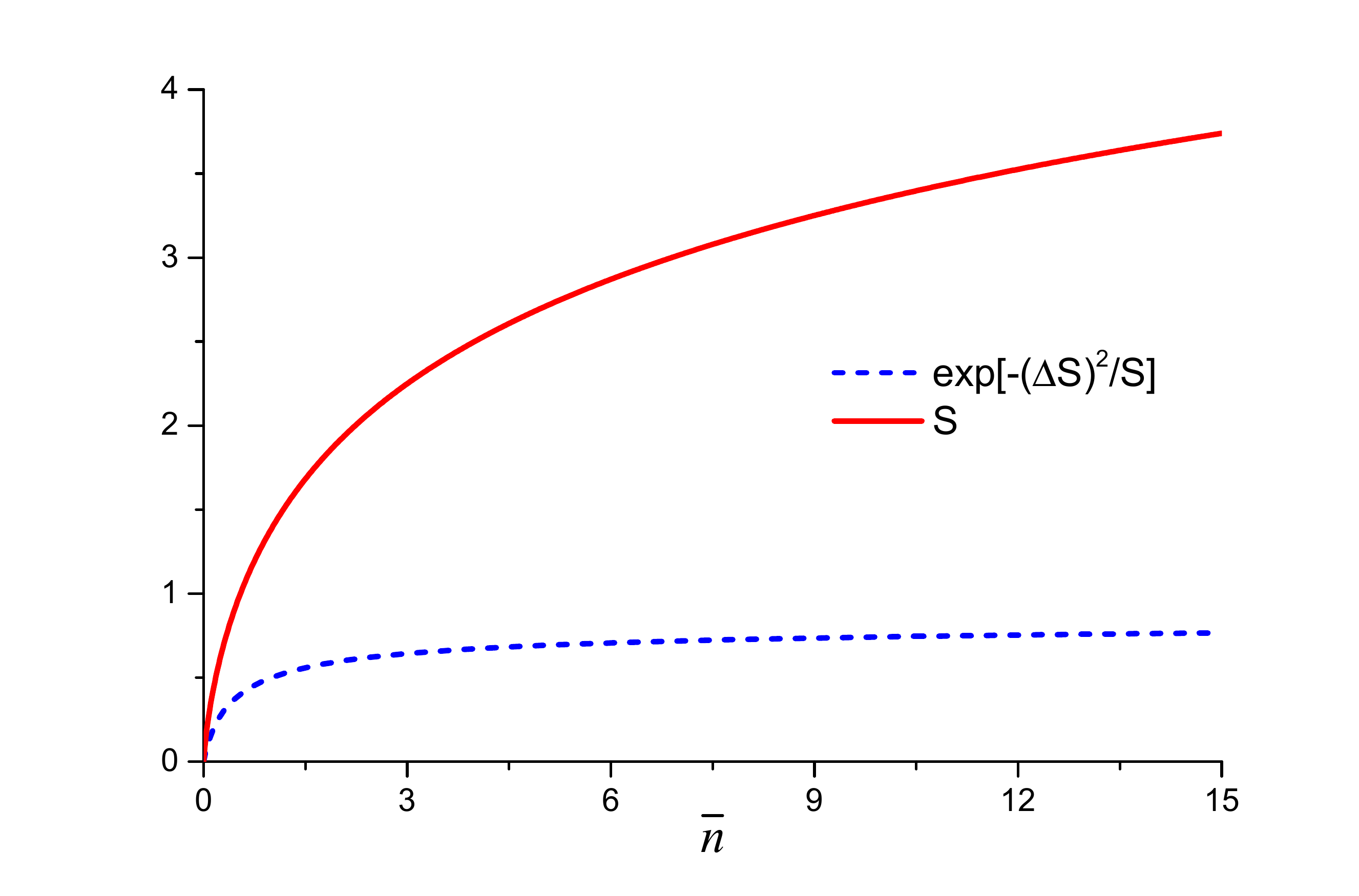} 
\caption{We plot the entropy, $S$, and the MP, $e^{-(\Delta S)^{2}/S}$ for the thermal distribution (\ref{Th1}) as a function of $\bar{n}$.}
\label{SDSTh}
\end{figure}
\section{MP in the Jaynes-Cummings model}
The interaction Hamiltonian for the atom-field interaction in the rotating wave approximation, i.e., for the Jaynes--Cummings model is given by \cite{JCM,Gerry}
\begin{equation}
\label{SM1}
\hat{H}_{I}=\lambda\left(\hat{a}^{\dagger}\sigma_{-}+\hat{a}\sigma_{+}\right)\,,
\end{equation}
where we have considered  the two-level atom transition frequency equal to the  quantized field frequency, i.e. the resonant interaction. Here $\hat{a}$ and $a^{\dagger}$ are the field annihilation and creation operators, respectively, and $\sigma_{+}=|e\rangle \langle g|$ and $\sigma_{-}=|g\rangle \langle e| $ are the rising and lowering operators for the atom (Pauli spin matrices). The kets $|e\rangle$ and $|g\rangle$  represent the excited and ground states of the atom, respectively. The parameter $\lambda$ is the coupling strength, also called the Rabi frequency. It is direct to obtain the evolution operator for the Hamiltonian above \cite{Stenholm}, which may be written as
\begin{equation}
\label{SM2}
\hat{U}_{I}=
\left(\begin{array}{ll}
\cos\left(\lambda t\sqrt{\hat{a}\hat{a}^{\dagger}}\right) & - i  \, \hat{V}\sin \left(\lambda t \sqrt{\hat{a}^{\dagger}\hat{a}} \right)\vspace{0.15cm} \\
- i \,  \hat{V}^{\dagger}\sin \left(\lambda t \sqrt{\hat{a}\hat{a}^{\dagger}} \right) &\cos\left(\lambda t\sqrt{\hat{a}^{\dagger}\hat{a}}\right)
\end{array}\right)\,,
\end{equation}
where  $\hat{V}=\frac{1}{\sqrt{\hat{a}\hat{a}^{\dagger}}}\hat{a}$ and $\hat{V}^{\dagger}=\hat{a}^{\dagger}\frac{1}{\sqrt{\hat{a}\hat{a}^{\dagger}}}$ are the London operators \cite{London} also known as Susskind-Glogower operators \cite{Susskind}. The evolved wavefunction for an initial state given by the initial quantized  field in  coherent state \cite{Glauber}, $\ket{\alpha}$, and the atom in its excited state  is given by
\begin{equation}
\label{SM14}
\ket{\psi}=\ket{\psi_{1}}\ket{e}+\ket{\psi_{2}}\ket{g}\,,
\end{equation}
with the unnormalized wavefunctions 
\begin{equation}
\label{SM15}
\ket{\psi_{1}}=\cos\left(\lambda t\sqrt{\hat{n}+1}\right)\ket{\alpha}\,, \,\,  
\ket{\psi_{2}}=-i\hat{V}^{\dagger}\sin\left(\lambda t\sqrt{\hat{a}\hat{a}^{\dagger}}\right)\ket{\alpha}\,.
\end{equation}
\begin{figure}[h]
\centering
\includegraphics[width=12cm]{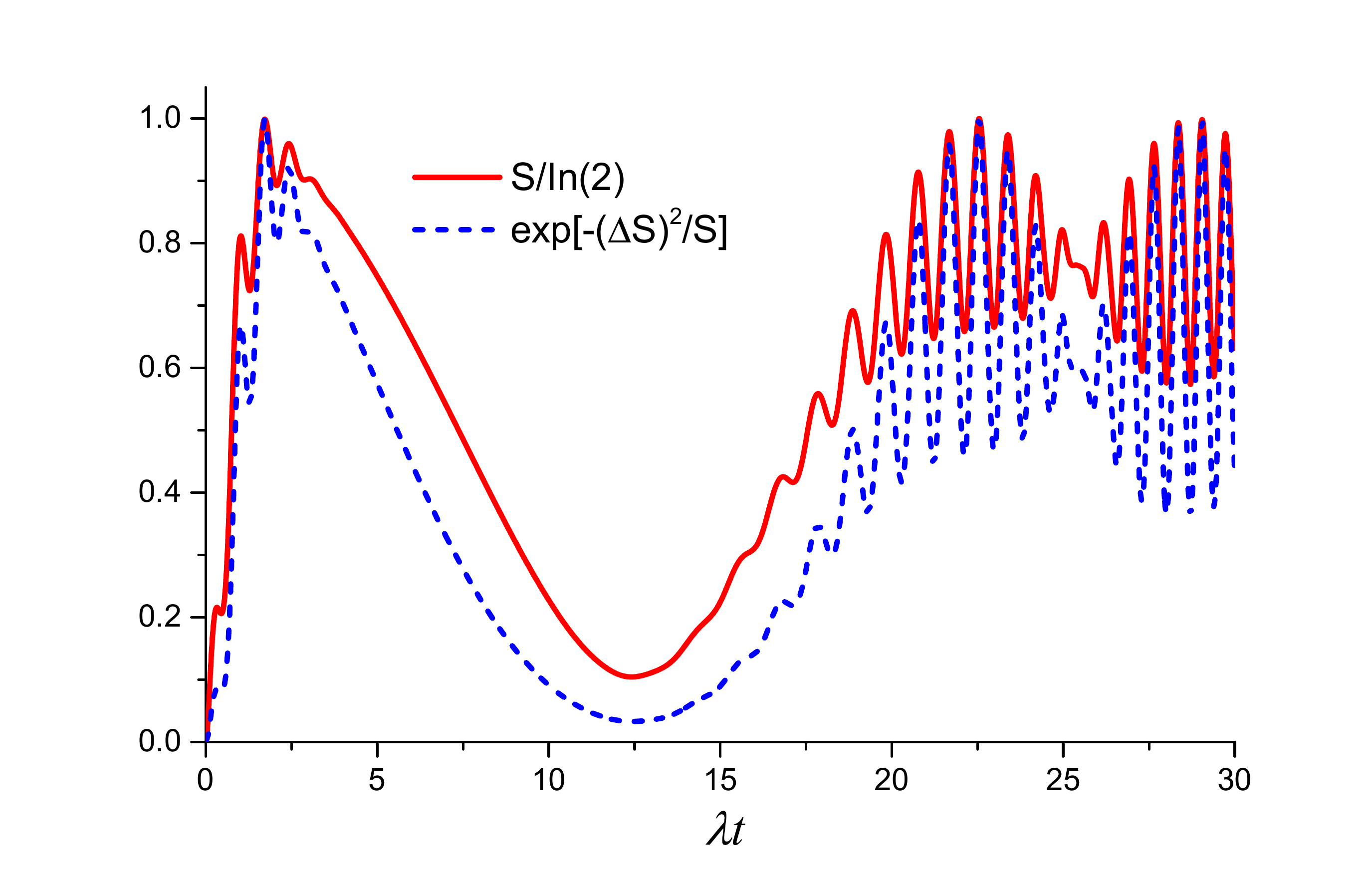} 
\caption{ We plot the atomic entropy, $S/\ln2$ and the atomic MP, $e^{-(\Delta S)^{2}/S}$, as  a function of $\lambda t$ for an initial coherent state for the field with $\alpha=4$, and the atom in its excited state $\ket{e}$.}
\label{SDSJCM}
\end{figure}
From these equations we may find the atomic and field density matrices as
\begin{equation}
\label{SM16}
\hat{\rho}_{_{A}}=
\left(\begin{array}{ll}
\braket{\psi_{1}|\psi_{1}}& \braket{\psi_{1}|\psi_{2}}^{*}\vspace{0.15cm}\\
\braket{\psi_{1}|\psi_{2}}& \braket{\psi_{2}|\psi_{2}} 
\end{array}\right),
\end{equation}
and
\begin{equation}
\label{SM17}
\hat{\rho}_{_{F}}=\ket{\psi_{1}}\bra{\psi_{1}}+\ket{\psi_{2}}\bra{\psi_{2}}.
\end{equation}
Equation (1) tells us that the entropy for the field and the atom are equal because the total entropy for the state (28) is zero. Then we find that the atomic entropy and the atomic MP are given by 
\begin{equation}
S_{A}=-\Lambda_{1}\ln\Lambda_{1}-\Lambda_{2}\ln\Lambda_{2}=S_{F}\,,
\qquad 
\label{SM21.2}
\Delta S_{A}=\sqrt{\Lambda_{1}\ln^{2}\Lambda_{1}+\Lambda_{1}\ln^{2}\Lambda_{2}-S^{2}_{A}}=\Delta S_{F}\,,
\end{equation}
where $\Lambda_1$ and  $\Lambda_{2}$ are the eigenvalues of the atomic density matrix (30).

In Figure 5 we plot the normalized  entropy and the  MP for the atom. Again, as in former cases, both follow similar behaviour, showing maximums at exactly the same interaction times.  We should stress again that we needed {\it a priori} information in order to know how to normalize the entropy.

 {
\section{Damped wave packet}
We now turn our attention to the problem of a decaying field, that may be described by a master equation \cite{Carmichael}. From Phoenix  \cite{Phoenix2} we know that the evolution of a quantized field, initially in a superoposition of coherent states, subject to decay is given by
\begin{eqnarray}
\label{E1} \nonumber
\hat{\rho}(t) &=& N\left(\ket{\alpha e^{-\gamma t/2}}\bra{\alpha e^{-\gamma t/2}} + \braket{\beta|\alpha}^{1-\exp(-\gamma t)}\ket{\alpha e^{-\gamma t/2}}\bra{\beta e^{-\gamma t/2}}\right) \\
&+&\left( \braket{\alpha|\beta}^{1-\exp(-\gamma t)}\ket{\beta e^{-\gamma t/2}}\bra{\alpha e^{-\gamma t/2}} + \ket{\beta e^{-\gamma t/2}}\bra{\beta e^{-\gamma t/2}} \right)\,, 
\end{eqnarray}
where 
\begin{equation}
\label{E2}
N = \frac{1}{2+2\mathrm{Re}\braket{\alpha|\beta}}\,,
\end{equation}
and $\gamma$ is the rate at which the field decays.
The density operator (\ref{E1}) may be rewritten as
\begin{equation}
\label{E3}
\hat{\rho}(t) = \ket{\psi_{1}}\bra{\psi_{1}} + \ket{\psi_{2}}\bra{\psi_{2}} \,,
\end{equation}
with
\begin{eqnarray}
\label{E4}
\ket{\psi_{1}} &=& \sqrt{N}\left(\ket{\alpha e^{-\gamma t/2}} + \braket{\alpha|\beta}^{1-\exp(-\gamma t)}\ket{\beta e^{-\gamma t /2}}\right) \,, \nonumber\\
\ket{\psi_{2}} &=& \sqrt{N\left(1-|\braket{\alpha|\beta}|^{2(1-\exp(-\gamma t))}\right)}\ket{\beta e^{-\gamma t /2}} \,,
\end{eqnarray}
from where the von Neumann entropy and its fluctuations are given by
\begin{equation}
\label{E5}
S = - \lambda_{+}\ln \lambda_{+} - \lambda_{-} \ln \lambda_{-} \,,
\qquad
\Delta S = \sqrt{\lambda_{+}\lambda_{-}}\left|\ln \left(\frac{\lambda_{+}}{\lambda_{-}}\right)\right|,
\end{equation}
where the eigenvalues are
\begin{equation}
\label{E7}
\lambda_{\pm} = \frac{1}{2}\pm \frac{1}{2}\sqrt{(\mathrm{P}_{11}-\mathrm{P}_{22})^{2}+4|\mathrm{P}_{12}|^{2}} \,,
\end{equation}
with
\begin{eqnarray}
\label{E8}
\mathrm{P}_{11} &=& N\left(1+2\mathrm{Re}\braket{\alpha|\beta} + |\braket{\alpha|\beta}|^{2(1-\exp(-\gamma t))}\right) \,, \nonumber\\
\mathrm{P}_{12} &=& N\left(\braket{\alpha|\beta}^{\exp(-\gamma t)}+ \braket{\beta|\alpha}^{1-\exp(-\gamma t)}\right)\sqrt{\left(1- |\braket{\alpha|\beta}|^{2(1-\exp(-\gamma t))}\right)}\,, \nonumber\\
\mathrm{P}_{22} &=& N\left(1- |\braket{\alpha|\beta}|^{2(1-\exp(-\gamma t))}\right) \,.
\end{eqnarray}
In Figure 6 we plot the MP and the normalized entropy for the damped wave packet initially in a superposition of coherent states. It may be seen the same behaviour, although, the von Neumann entropy needs to be normalized while the MP does not.  }
\begin{figure}[h!] 
\centering
\includegraphics[width=12cm]{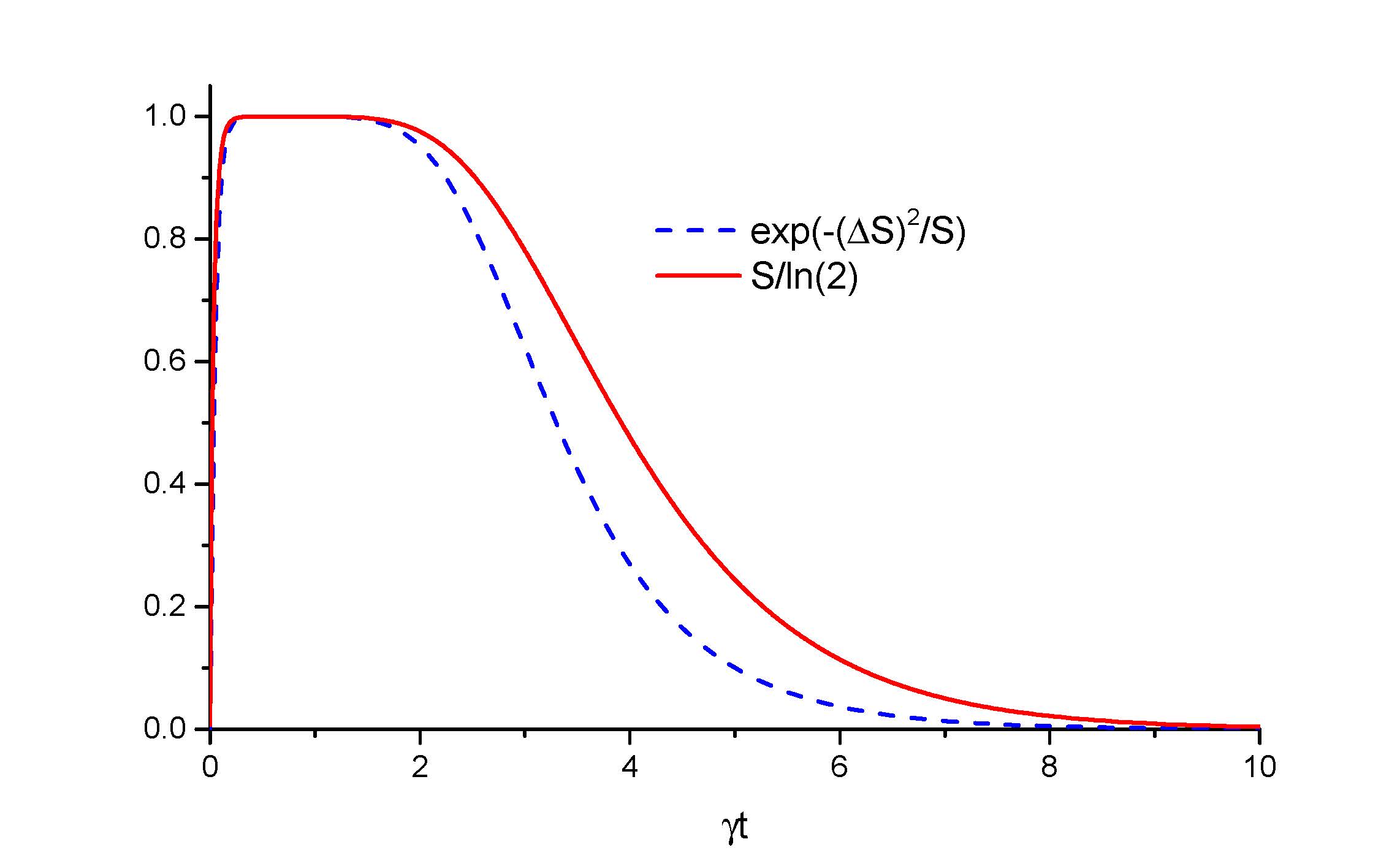} 
\caption{This plot shows the time evolution of the normalized  entropy, $S/\ln 2$ (dashed line), and the MP,  $Q_{S}$ (solid line), for $\alpha = 2$ and $\beta = 7$.}
\label{SF}
\end{figure}
\section{Conclusion}
We have introduced a new quantifier for mixedness of a given state based on entropy fluctuations. This parameter has the property that it is bounded from zero, for pure states, to one for completely mixed states. As we have shown, for quantized fields, entropy requires to have {\it a priori} knowledge about the state to properly normalize the entropy while for  the MP it is not necessary such knowledge.  {The mixedness parameter provides us with information about that either we have a completely mixed state or a reduction of the Hilbert space has occurred.} We could not properly normalize the entropy in the example given for the thermal distribution and therefore could not bound it to one.

 {In order to be clearer, we give the following example: consider a system living in a five-dimensional Hilbert space and described by the density matrix
\begin{equation}
\label{example}
\hat{\rho}_{5\times 5}^{(1)}=\frac{1}{5}(|1\rangle\langle 1|+|2\rangle\langle 2|+\dots |5\rangle\langle 5|),
\end{equation}
clearly, this is a maximally mixed density matrix, and the linear or von Neumann entropy would show so.  The  parameter we introduced, would also be maximum, i.e., $Q_s^{(1)}=1$.
If we consider, instead the density matrix
\begin{equation}
\label{example2}
\hat{\rho}_{5\times 5}^{(2)}=\frac{1}{2}(|1\rangle\langle 1|+ |5\rangle\langle 5|),
\end{equation}
that would have a linear entropy \cite{Peters2004b} $\xi_{5\times 5}^{(2)}=\frac{5}{4}(1-Tr\{(\hat{\rho}_{5\times 5}^{(2)})^2\}=\frac{5}{8}$, which would not be a maximum, while the parameter we introduced would attain its maximum value. However, we could give a third example, $
\hat{\rho}_{5\times 5}^{(3)}$ that presents some coherences between the different states and  for which $\xi_{5\times 5}^{(3)}=\frac{5}{8}=\xi_{5\times 5}^{(2)}$, but the mixedness parameters would not be equal,  $Q_s^{(2)}\ne Q_s^{(3)}$. Therefore there is some information in the MP no present in the linear entropy that indicates that the Hilbert space has been reduced and, in the reduced Hilbert space,  the state is a maximum mixed state.}

\bigskip
{\bf Author Contributions}: Jorge A. Anaya-Contreras conceived the idea and developed it under Arturo Z\'u\~niga-Segundo and H\'ector M. Moya-Cessa supervision. The manuscript was written by all authors, who have read and approved the final manuscript.
\bigskip

{\bf Conflicts of Interest}: The authors declare no conflict of interest.

\end{document}